\documentstyle[mprocl]{article}
\input{psfig}
\begin{document}

\title{\Large Current--Carrying Cosmic String Loops Leading to Vortons}
\author{\large Alejandro GANGUI}
\address{\large DARC -- Observatoire de Paris--Meudon, 92195 Meudon, France}
\author{\large Edgard GUNZIG}
\address{\large Universit\'e Libre de Bruxelles, CP231 1050 Bruxelles, Belgium}
\maketitle\abstracts{\large \sl 
In this article we review recent work aimed at showing 
explicitly the influence of electromagnetic self corrections 
on the dynamics of a circular vortex line endowed with a current at 
first order in the coupling between the current and the
self--generated EM-field.}  

\vspace{1cm}
\begin{center}
{\sl To appear in \\
Proceedings of the Eighth Marcel Grossmann Meeting \\
on General Relativity, Gravitation and Relativistic Field Theories. \\
22--27 June 1997, Jerusalem}
\end{center}

\newpage

\title{Current--Carrying Cosmic String Loops Leading to Vortons}
\author{Alejandro GANGUI}
\address{
DARC -- Observatoire de Paris--Meudon, 92195 Meudon, France}
\author{Edgard GUNZIG}
\address{Universit\'e Libre de Bruxelles, CP231 1050 Bruxelles, Belgium}
\maketitle\abstracts{
In this article we review recent work aimed at showing 
explicitly the influence of electromagnetic self corrections 
on the dynamics of a circular vortex line endowed with a current at 
first order in the coupling between the current and the
self--generated EM-field.}  
Topological defects are widespread in most extensions of the standard
model of particle interactions. Among them cosmic strings could have
been produced as vacuum vortex lines as the outcome of early 
phase transitions.
A resulting network of 
strings would indeed have key features 
and leave unique traces in a variety of present day observables,
ranging from the small--scale anisotropies in the CMB 
radiation to the lensing of distant astrophysical sources. 
Extensions of the simplest models of cosmic strings, as the one
envisaged by Witten, 
involve extra degrees of freedom which are
coupled to the vortex--forming Higgs field. 
This is the source of currents, bound to the core of the vortices,
that may play a fundamental role in the dynamics of the defects and, 
in the case of string loops that we will consider here, may build up so
strong as to compensate the natural string tension towards collapse. 
This observation, first pointed out by Davis and Shellard in 1989, led 
to the study of rotating, equilibrium particle--like {\it vorton}
configurations and to the eventual excess danger implied by these 
remnants for the standard cosmology. 

Regarding vorton formation, not any arbitrary cosmic string loop, with
given characteristic ``quantum'' numbers, will in general end up
as a vorton. 
The way in which one can quantify the analysis is by studying an 
initial configuration (circular in our present work) by means of the
covariant macroscopic formalism for general strings developed by
Carter, 
and letting this initial state evolve in
order to see whether it attains stability or not.
This project is presently in progress and results were already
obtained both for the neutral current--carrying case 
and also when the effect
of including electromagnetic (e.m.) self coupling is taken into account.
Radiative effects are of course important for the complete analysis 
and are currently under investigation. 
Let us recapitulate briefly the microphysics setting and its connection 
with the macroscopic string description we use in our analysis. 
We consider a Witten--type bosonic superconductivity
model in which the fundamental Lagrangian is invariant under the
action of a U(1)$\times$U(1) symmetry group. The first U(1) is
spontaneously broken through the usual Higgs mechanism in which
the Higgs field $\Phi$ acquires a non--vanishing vacuum expectation value.
Hence, at an energy scale $\sim m$ we are left with a network of
ordinary cosmic strings with tension and energy per unit length 
$T \sim U \sim m^2$, as dictated by the Kibble mechanism. 
The Higgs field is coupled not only with its associated gauge vector 
but also with a second charged scalar boson $\Sigma$, the {\it current
carrier} field, which in turn obeys a quartic potential.
A second phase transition breaks the second U(1) gauge (or global, in
the case of neutral currents) group and, at an energy scale $\sim m_*$,
the generation of a current--carrying condensate in the vortex makes
the tension no longer constant, but dependent on the magnitude of the
current, with the general feature that $T \leq m^2 \leq U$, breaking
therefore the degeneracy of the Nambu--Goto strings.
The fact that $|\Sigma|\neq 0$ in the string results in that either 
electromagnetism (in the case that the associated gauge vector $A_\mu$
is the e.m. potential) or the global U(1) is spontaneously
broken in the core, with the resulting Goldstone bosons carrying
charge up and down the string.  
The invariance of the Lagrangian with respect to local changes in the
phase $\varphi$ of the current carrier field 
$\Sigma = |\Sigma(x,y)|\exp[i\varphi(z,t)]$
is expressed by the conservation of the Noether current 
$2|\Sigma|^2(\partial^\mu\varphi-eA^\mu)$, where we implicitly consider
a string along the $z$ direction and where the physically relevant
scalar boson amplitude cannot depend on the internal string
coordinates, $a= z, t$. 
In order to get the total macroscopic current one just needs to
integrate over the string cross section (taking $A_\mu\sim {\rm
const}$ in the core) to get the current
$z_a = 2\tilde\Sigma(\partial_a\varphi-eA_a) \equiv
2\tilde\Sigma\varphi_{|a}$, where 
$\tilde\Sigma=\int_{\rm core}dx dy |\Sigma|^2$.
In the macroscopic string description a key r\^ole is played by the squared of
the gradient of $\varphi$ in characterizing the local state of the string
through $w =\gamma^{ab}\varphi_{|a }\varphi_{|b}$.
The dynamics of the system is determined by the Lagrangian 
${\cal L}\{ w \}$. From it we get the conserved particle current vector
$z_a=-\partial{\cal L}/\partial\varphi^{|a} = {\cal
K}^{-1}\varphi_{|a} = 2 \tilde\Sigma \varphi_{|a}$, where we define 
${\cal K}^{-1}\equiv -2 d{\cal L} / dw$ which is in turn proportional
to the amplitude of the condensate $\tilde\Sigma$.
From the above middle equality we get 
${\cal L}\{ w \} = -m^2-\varphi_{|a}\varphi^{|a} / 2{\cal K} = 
-m^2-w/2{\cal K}$ which, for weak currents (small $w$) coincides 
(recall ${\cal K}\to 1$ for $w\to 0$) with the generalization of the
Nambu--Goto model given by Carter and Peter in 1995: 
${\cal L}\{ w\} = -m^2 - {1\over 2} m_*^2 \ln \left\{ 1 + {w / m_*^2}
\right\} .$

The description of the dynamics of macroscopic strings in terms of
${\cal L}\{ w\}$ is easily understandable, given the clear physical 
meaning of the phase $\varphi$. 
However, in what follows, an equally powerful dual formalism will 
be used, it being based on the master function 
$\Lambda \{ \chi \}$, with $w={\cal K}^2\chi$ and 
$\chi = z^\mu z_\mu$, where $z^\mu$ is the tangential current vector
{\it on} the string worldsheet defined by $z^\mu = z^a x^\mu_{,a}$ in terms
of the current $z^a$ {\it in} the worldsheet. 
Both 
$\Lambda$ and 
${\cal L}$ are related by a Legendre 
transformation 
$\Lambda={\cal L}+{\cal K}\chi$. 
These functions provide values for the string $U$ and $T$
depending on the signs of the state parameters $\chi$ and $w$.
Hence we have $U=-\Lambda$ ($U=-{\cal L}$), $T=-{\cal L}$ ($T=-\Lambda$)
for timelike (spacelike) currents $\chi < 0$ ($\chi > 0$). 
The validity of this description follows from the requirement of
local stability which demands that the squared
speeds $c_{_{\rm E}}^{\, 2}=T/U$ and $c_{_{\rm L}}^{\,2} =-dT/dU$ of
extrinsic 
and longitudinal 
perturbations be positive. Therefore 
${\cal L} /\Lambda>0>d{\cal L} / d\Lambda$ in both regimes.

Electromagnetic corrections have
recently been calculated by Carter (these proceedings). 
The result is surprisingly simple to implement. 
The correction enters through a modification of the 
string equation of state. 
The regularization of the divergent $A_\mu$ 
leads to a renormalization of $\Lambda$ of the kind
$\Lambda \to \Lambda + {1\over 2} \lambda q^2 \chi$, where $\lambda$
is given by the string current self generated potential
$A^\mu = \lambda j^\mu$, with the e.m. current 
$j^\mu = q z^\mu$ flowing along the string and 
$\lambda = 2 \ln (m_\sigma \Delta )$, where $\Delta$ is an infrared
cutoff scale to compensate for the
asymptotically logarithmic behavior of the e.m. 
potential and $m_\sigma$ the ultraviolet cutoff
corresponding to the effectively finite thickness of the charge
condensate, i.e., the Compton wavelength of the current-carrier
$m_\sigma^{-1}$.
In the practical situation of a
closed loop, $\Delta$ should at most be taken as the total length of
the loop.
\begin{figure}
        \begin{flushleft}
        \mbox{
                \psfig{file=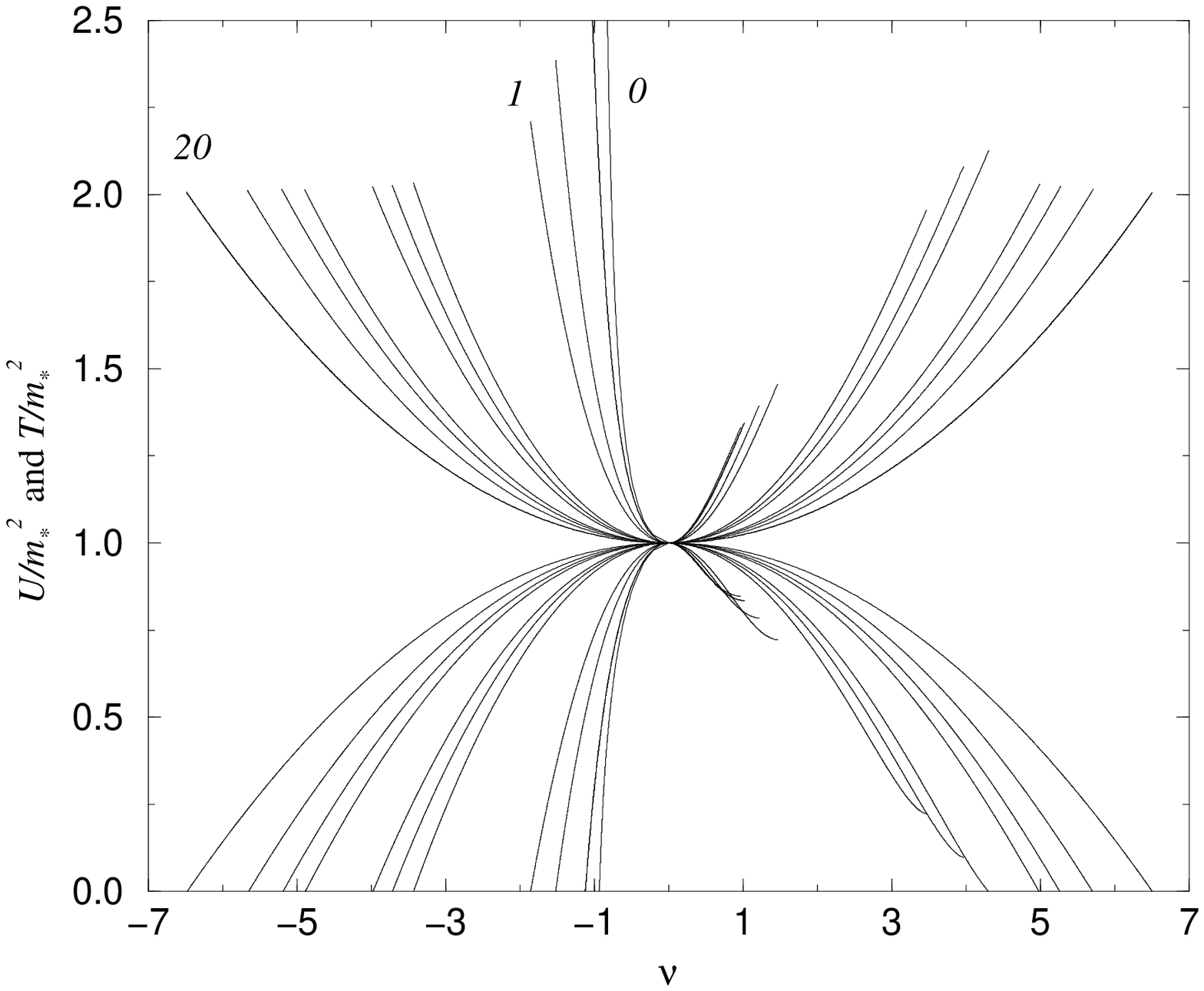,width=6.5cm}\hspace{-1cm}
                \psfig{file=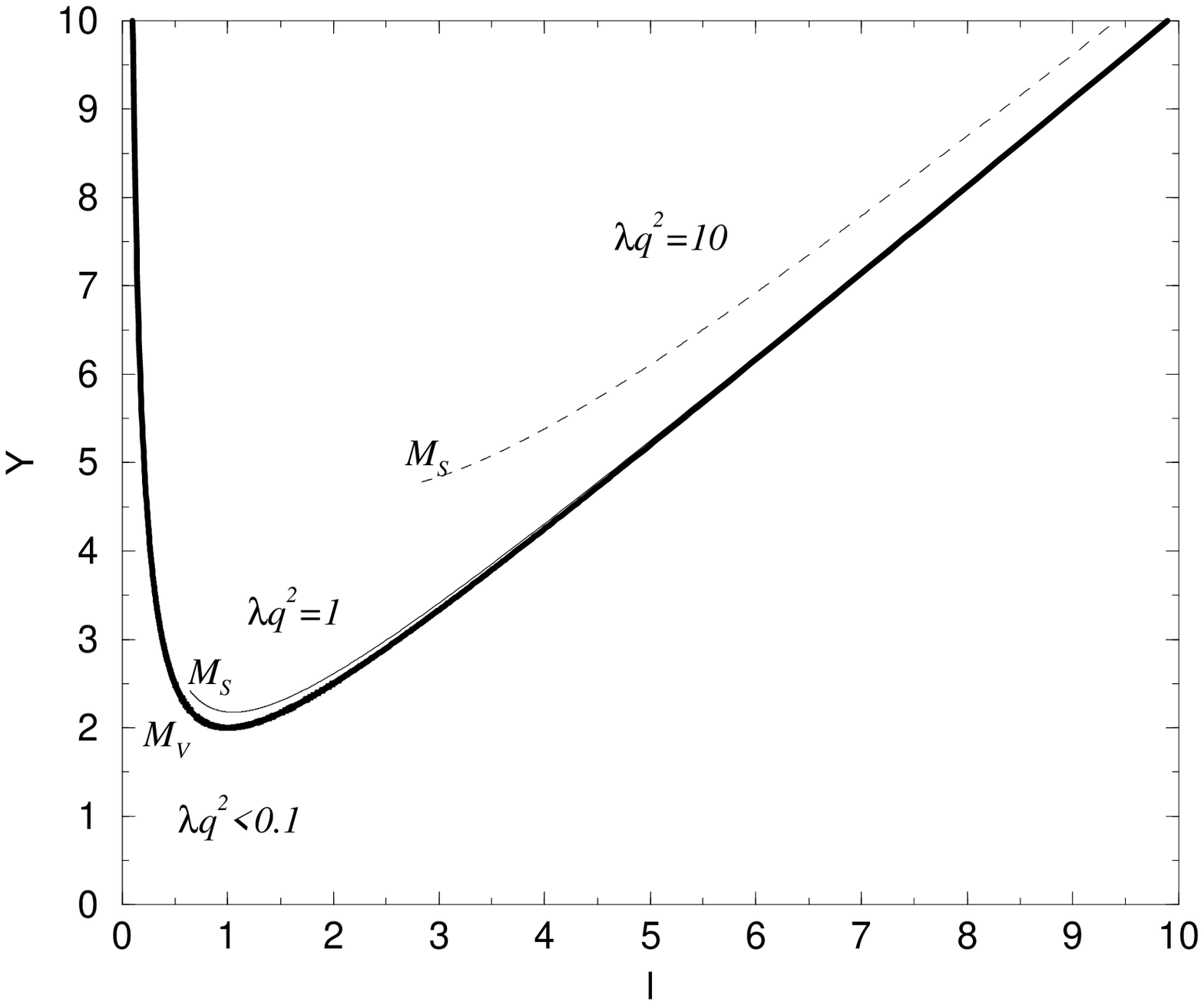,width=6.5cm}
             }
        \end{flushleft}
\vspace{-1cm}
\caption{\label{fig4}
{\small 
Left panel: $U$ (upper set of curves) and $T$ (lower set of curves) 
versus $\nu$. 
Increasing values of $\lambda q^2$ enlarges the corresponding curve. 
Right panel: Variations of the self potential $\Upsilon$. 
Three curves are shown showing the qualitatively different cases in
which the loop either reaches stability (thick curve) or eventually decays 
(light curves). 
}}
\vspace{-0.6cm}
\end{figure}
We are now able to compute the variation of the string equation of state
with the e.m. self correction $\lambda q^2$, which we plot
in Fig 1 (left panel) as a function of the sign--preserving square root of the
state parameter $\nu = $~Sign$(w) \sqrt{|w|}$. It is interesting to
see that the inclusion of self correction allows larger currents along
the string before reaching the saturation point.

Now, regarding the motion of circular vortex rings in flat space, a
previous analysis of one of us 
shows that the variation of the string radius follows from the
equation
$M\sqrt{1-\dot r^2} = \Upsilon (r)$, where $M$ is the string's 
total mass and $\Upsilon (r)$ is the self potential on which the
string's radius evolves. $\Upsilon$ itself can be written in terms of 
conserved quantities such as the number of carrier particles in the
loop $Z$ and the topologically conserved (in the 2D
worldsheet) winding number $N$ of the carrier--field's phase $\varphi$ 
around the loop. 
It is an easy task to derive the form of the ring radius in terms of these
conserved quantities, $r^2 \propto Z^2 (b^2-{\cal K}^2) / {\cal K}^2 \chi$,
where $b\equiv |N/Z|$. This tells us that the nature of
the current (whether $\chi$ is positive or negative) 
depends on the sign of $b^2-{\cal K}^2$, where $b$ characterizes a
particular current state of the string and ${\cal K}$ is given by the
particular macroscopic model (through its Lagrangian) including
e.m. self corrections.
What one finds from local stability considerations is that the range
of variation of ${\cal K}$ is
$\lambda q^2 \leq {\cal K} \leq 1+\lambda q^2$ for $\chi \leq 0$,
whereas $1+\lambda q^2 \leq {\cal K} \leq 2+\lambda q^2$ in the 
$\chi \geq 0$ case.
Therefore, it is only possible for $\chi$ to be positive if $b\geq 1+
\lambda q^2$, and negative otherwise. 
As the critical value $b_c=1+\lambda q^2$ was unity in the decoupled 
case, we see that interestingly enough e.m. corrections 
can modify the nature of the current for a given set of $Z$ and $N$.

We plot the variations of the self potential $\Upsilon$ with the
ring's circumference $\ell$ and the e.m. self 
coupling $\lambda q^2$ for $(m/m_*)^2=1=b$ in the right panel of Fig 1.
The thick curve, a `safe' stable zone in parameter space is shown. 
For all regimes, this zone is 
limited to $\lambda q ^2 \leq b \leq \lambda q^2 + 2$,
a condition which is increasingly restrictive as $\lambda q^2$
increases, and may even forbid vorton formation altogether for a very
large coupling, with interesting consequences for the vorton excess problem.

\noindent {\sl Extensive list of references is given in papers 
{\tt hep-ph/9609401} and {\tt hep-ph/9705204}.
A.G. thanks the British Council for partial finantial support.
This work was partially supported by EEC grants Nr 
PSS*0992 and CI1*CT94-0004.
We would like to thank C.Boehm, B.Carter and P.Peter 
for enlightening discussions.}

\end{document}